\documentclass[
    ,final            
  ]
  {aipproc}

\layoutstyle{8x11double}
\usepackage{natbib}

\begin{document}

\title{Cepheids and Long Period Variables in M\,33}


\author{Anne Pellerin}{
  address={George P. and Cynthia W. Mitchell Institute for Fundamental Physics and Astronomy, \\ Department of Physics, Texas A\&M University, College Station, TX 77843, USA}
}

\author{Lucas M. Macri}{
  address={George P. and Cynthia W. Mitchell Institute for Fundamental Physics and Astronomy, \\ Department of Physics, Texas A\&M University, College Station, TX 77843, USA}
}

\author{Andrew K. Bradshaw}{
  address={George P. and Cynthia W. Mitchell Institute for Fundamental Physics and Astronomy, \\ Department of Physics, Texas A\&M University, College Station, TX 77843, USA}
}
\author{Krzysztof Z. Stanek}{
 address={Department of Astronomy, The Ohio State University, 140 West 18th Avenue, Columbus, OH 43210, USA}
}

\begin{abstract}
We are conducting a long-term photometric survey of the nearby galaxy M\,33 to discover Cepheids, eclipsing binaries, and long-period variables. The dataset combines previously-obtained optical images from the DIRECT project with new observations acquired at the WIYN 3.5m telescope. The entire data set spans over 7 years with excellent synoptic coverage which will enable the discovery and characterization of stars displaying variability over a wide range of timescales (days, weeks, months, years).
 
In this preliminary work we show representative light curves of different variables we found so far in two fields, color-magnitude diagrams, and optical Cepheid Period-Luminosity relations for M\,33. The ultimate goal of the project is to provide an absolute calibration of the Cepheid Period-Luminosity relation, and to study its metallicity dependence at optical wavelengths.

\end{abstract}

\maketitle


\section{The Project}

Cepheid variables are a fundamental distance indicator that can be applied out to $\sim$40\,Mpc with current instrumentation. Our project aims to provide an absolute calibration and a determination of the impact of metallicity on the Cepheid Period-Luminosity (PL) relation, also known as the Leavitt Law. Approximately 900 BVI images of the Local Group galaxy M\,33 were obtained between 1996 to 2001 by the DIRECT project, mainly using the F.~L. Whipple Observatory 1.2m telescope. Another $\sim$1000 BVI images were obtained in 2002-2003 with the WIYN 3.5m telescope to extend the time coverage and provide higher spatial resolution. The full dataset covers the whole disk and spans about 7.4 years. This will enable us to identify Cepheids across the disk of M\,33, which is known to display a large metallicity gradient \citep{rowe05,mag07}. With such a long time coverage we also expect to discover many other variable stars, including long period variables.

In this contribution, we present a preliminary analysis of two fields. Representative light curves of Cepheids and long-period variables as well as the color-magnitude diagrams and PL relations for those fields are shown. 

\section{Data and Photometric Analysis}

We analyzed BVI ground-based images that cover two fields in M\,33. The first field is a 10$' \times$5$'$ field centered on the spiral arm located north of the nucleus (hereafter m0b field). The second is a 11.5$'\times$11.5$'$ field west of the nucleus centered on the western arm (hereafter Y3ZC field). The Y3ZC dataset originates from the DIRECT projects and spans 800 days. The m0b field dataset includes WIYN observations in addition to the DIRECT data and therefore covers over 2700 days. 

\begin{figure}
  \includegraphics[height=.356\textheight]{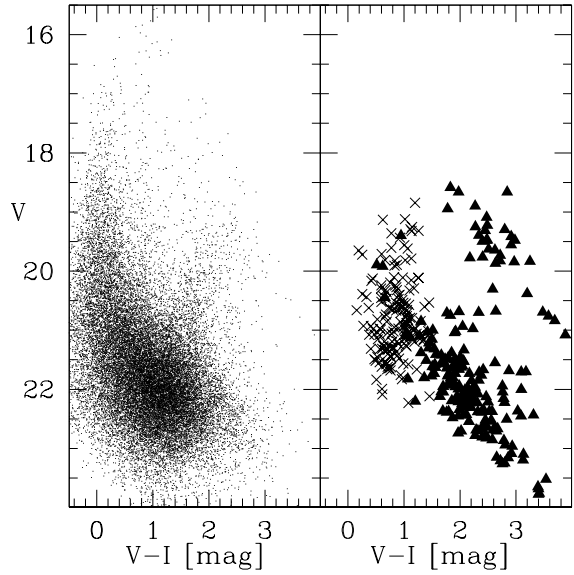}
  \caption{\label{cmd} Color-magnitude diagrams for the two fields combined. Left: all stars detected in the V- and I-bands. Right: variable stars only, with the Cepheids represented by crosses and long period variables ($>$ 150 days) by triangles. Long period variables with no V-band detection are not included.}
\end{figure}

PSF photometry was performed using DAOPHOT and ALLFRAME in the three bands \citep{stet94}. The photometric zeropoints and astrometry were determined using the M\,33 photometric catalog by \citet{mass06}. A color-magnitude diagram of the combined fields is presented in Fig.~\ref{cmd}. The variability index, J \citep{stet96}, was calculated to identify variable star candidates. We used the CLEAN algorithm \citep{rob87} to determine periods for all stars displaying a J-index larger than 0.75.  We then fit the observed V- and I-band light curves to the templates from \citet{stet96} to identify Cepheid stars. We then applied the same technique to the I-band light curves for stars with a J-index larger than 2.0 to find variable stars that were too red to produce a reliable light curve in the V-band.

\section{Preliminary Results}

\begin{figure}
  \includegraphics[height=.25\textheight]{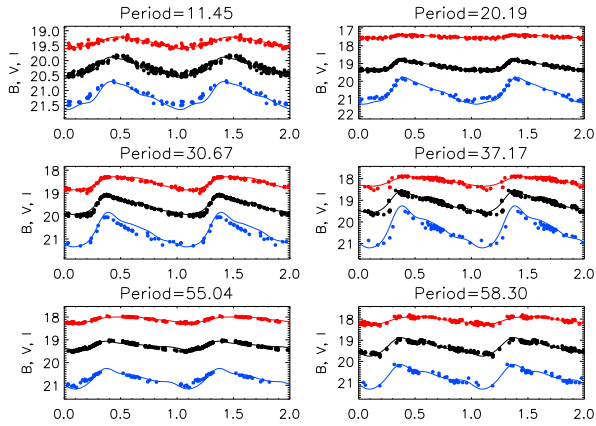}
  \caption{\label{cephlc} Examples of Cepheid light curves detected in the 11.5$' \times$11.5$'$ Y3ZC field, as a function of phase. For each panel the B-, V-, and I-bands are the lower, center, and upper curves respectively. The circles are the observations and the lines are the best-fit templates. Periods are in days.}
\end{figure}

We found more than 200 Cepheids and over 400 long period red variables thanks to the large time interval covered by our data. Proportionally speaking, we discovered about twice as many long period red variables in the smaller m0b field due to the extended time coverage provided by the additional WIYN data. Examples of Cepheid light curves detected in the Y3ZC field together with their best-fit template are presented in Fig.~\ref{cephlc}. 

\begin{figure}
  \includegraphics[height=.25\textheight]{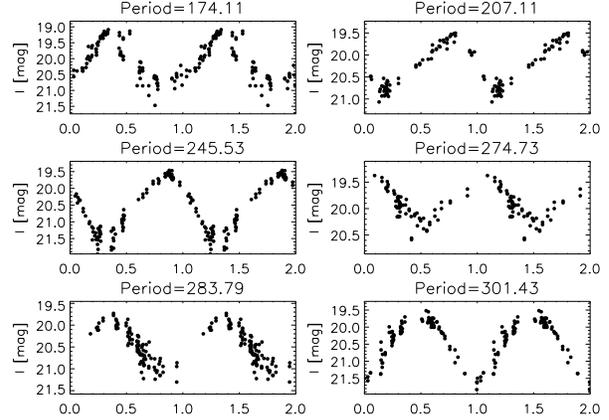}
  \caption{\label{mira} Examples of I-band light curves from long period variables detected in the m0b field, as a function of phase. Periods are in days.}
\end{figure}

\begin{figure}
  \includegraphics[height=.25\textheight]{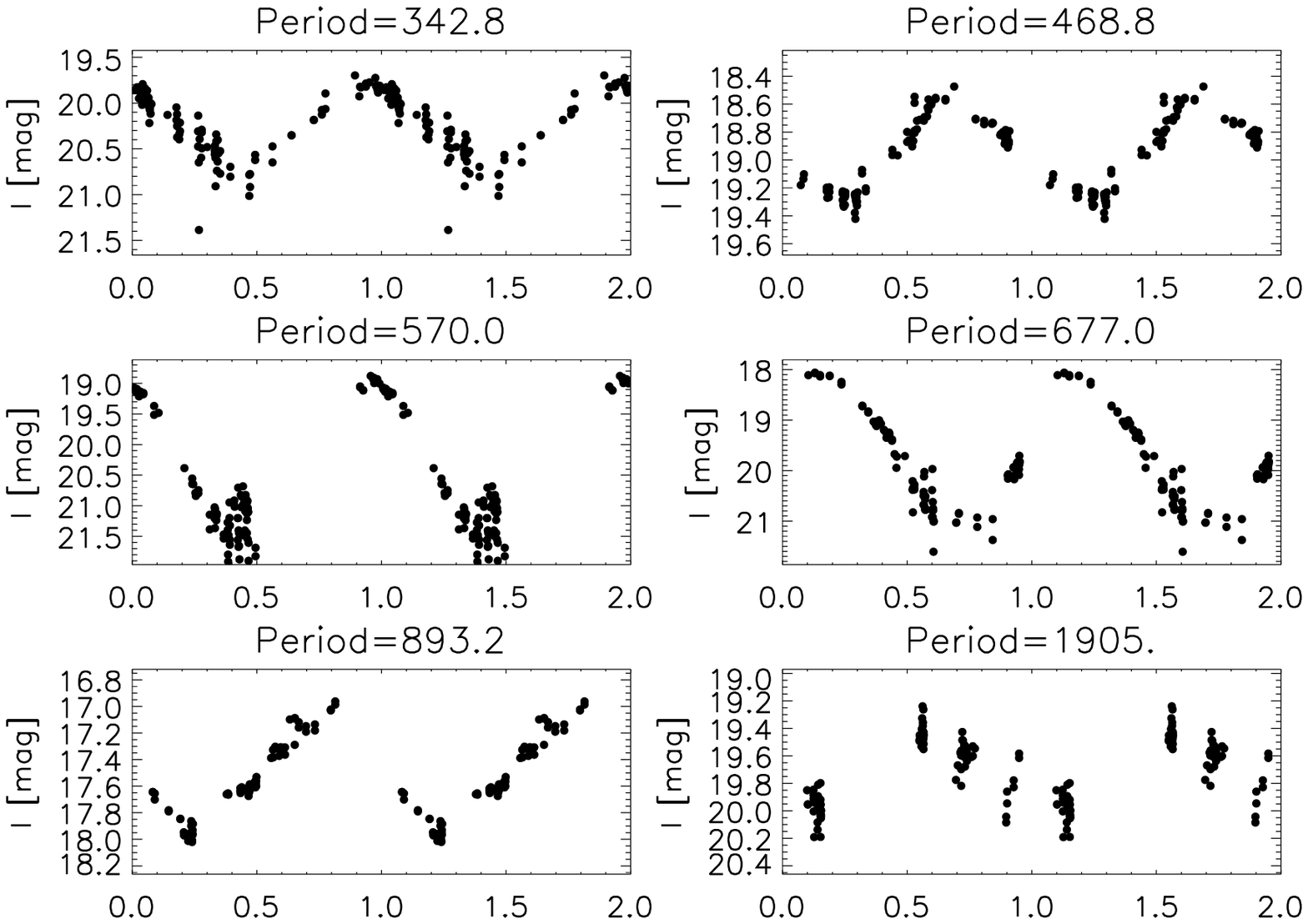}
  \caption{\label{ulp} Examples of I-band light curves from ultra long period variables detected in the m0b field, as a function of phase. Periods are in days.}
\end{figure}

\begin{figure}[!t]
  \includegraphics[height=.24\textheight]{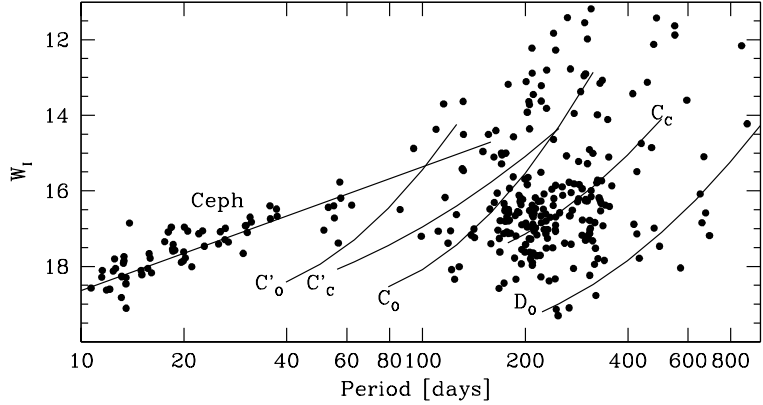}
  \caption{\label{wi} Wesenheit I index of variables detected in the I-band in both m0b and Y3ZC fields. The lines indicate locations of different types of variables as reported by \citet{sos07} for the LMC. C$_O$ and C$^\prime_O$: Oxygen-rich Miras and semi-regular PL-relations. C$_C$ and C$^\prime_C$: Carbon-rich Miras and semi-regular PL-relations. D$_O$: Oxygen-rich long secondary period PL-relation. Ceph: Cepheid PL-relation.}
\end{figure}

In the m0b field alone, we detected about 230 long period variables (P>100 days). A sample of their typical I-band light curves are shown in Figs.~\ref{mira} and \ref{ulp}. V- and B-band light curves were not found for many long period variables due to their extremely red colors. Most of them display an I-band luminosity variation of at least 1\,mag, sometimes reaching 2.5-3.0\,mag for the most extreme cases. Because of their location on the color-magnitude diagram (Fig.~\ref{cmd}), most of the long period variables are likely to be late-type stars. A significant fraction of them are seen on the red supergiant branch. 
The location of the long period variables on the Wesenheit dust-free PL relation (Fig.~\ref{wi}) also shows that they are in very good agreement with the carbon-rich and oxygen-rich PL relations for Miras and other semi-regular variables \citep{sos07}, which is consistent with their large variability of 1-3\,mag. In the future analysis of the whole dataset we expect to find many more long period red variables and study their PL relations in greater detail.

\begin{figure}
  \includegraphics[height=.35\textheight]{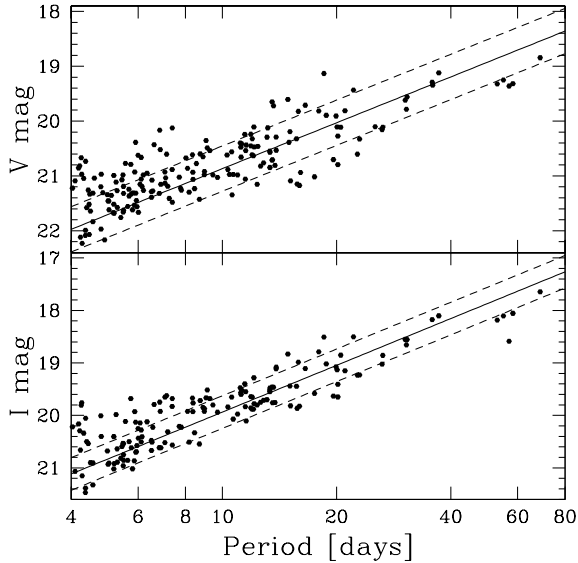}
  \caption{\label{plplot} Period-luminosity relation of Cepheids found in the two analyzed fields of M\,33. Dashed lines: 1$\sigma$ deviation.}
\end{figure}

We generated the Cepheid PL relations for both fields (Figs.~\ref{wi} and \ref{plplot}) and fit them with the PL relation from the OGLE survey for the LMC \citep{udal99}. The deviations below 8\,days are most likely due to blended objects and overtone plusators. Our preliminary distance modulus for M33 is consistent with \citet{mac01}. However, the analysis of the whole dataset  is needed to establish a more precise distance to M\,33. 

\section{Future Work}

The ultimate goal of this project is to study the PL relation dependence on metallicity by using Cepheids detected at various radii. The entire dataset is currently being analyzed in a similar way in order to gather a large sample of Cepheids and other long-period red variables.




\begin{theacknowledgments}
This work was supported by start-up funds from Texas A\&M University.
\end{theacknowledgments}



\bibliographystyle{aipproc}   

\bibliography{mybib}

\IfFileExists{\jobname.bbl}{}
 {\typeout{}
  \typeout{******************************************}
  \typeout{** Please run "bibtex \jobname" to optain}
  \typeout{** the bibliography and then re-run LaTeX}
  \typeout{** twice to fix the references!}
  \typeout{******************************************}
  \typeout{}
 }

\end{document}